\documentclass[a4paper,12pt,english]{article}

\pdfoutput=1

\usepackage[utf8x]{inputenc}	
\usepackage[english]{babel}
\usepackage[bindingoffset=0.3cm,textheight=22.5cm,hdivide={2.3cm,*,2.3cm},
vdivide={*,23cm,*}]{geometry}
\usepackage[svgnames]{xcolor}
\usepackage[pdftex,breaklinks,colorlinks=true,urlcolor=RoyalBlue,%
  linkcolor=blue,citecolor=blue]{hyperref}

\usepackage{amsmath,amsfonts,amssymb,slashed,graphicx,color,empheq}

\usepackage{cite}

\usepackage[toc,page]{appendix}

\def\nn{\nonumber}
\def\bea{\begin{eqnarray}}
\def\eea{\end{eqnarray}}

\newcommand{\eq}[1]{(\ref{#1})}

\def\a{\alpha}          
\def\b{\beta}           
    
\def\d{\delta}

\def\g{\gamma}

\def\l{\lambda} \def\L{\Lambda}

\def\s{\sigma}

\def\G{\Gamma}

\def\LP{\mathsf{P}}

  \def\cC{{\cal C}}
  
 \def\cH{{\cal H}} 
  
\def\cM{{\cal M}}  
 \def\cQ{{\cal Q}}

\def\cY{{\cal Y}}

\def\R{{\mathbb R}}

\def\N{{\mathbb N}}
\def\Z{{\mathbb Z}}
\def\one{\mbox{1 \kern-.59em {\rm l}}}

\def\({\left(}
\def\){\right)}
\def\diag{\mbox{diag}}
\def\Tr{{\rm Tr}}
\def\End{{\rm End}}

\def\mso{\mathfrak{so}}

\def\del{\partial}

\def\half{\frac{1}{2}}
\def\quarter{\frac{1}{4}}

\begin{document}

\makeatother



\renewcommand{\title}[1]{\vspace{10mm}\noindent{\Large{\bf

#1}}\vspace{8mm}} \newcommand{\authors}[1]{\noindent{\large

#1}\vspace{5mm}} \newcommand{\address}[1]{{\itshape #1\vspace{2mm}}}

\begin{titlepage}
\pdfbookmark[1]{Title}{Title}
\begin{flushright}
 UWThPh-2022-8 \\
\end{flushright}
\begin{center}

\title{Cosmic time evolution and propagator \\[1ex]
from a Yang-Mills matrix model}

\authors{Joanna L. Karczmarek\footnote{joanna@phas.ubc.ca}${}^\dagger$ and
Harold C. Steinacker\footnote{harold.steinacker@univie.ac.at}${}^\ddagger$}

\vskip 3mm

 \address{ 
${}^\dagger${\it Department of Physics and Astronomy, University of British Columbia,
6224 Agricultural Road, Vancouver, BC Canada V6T 1Z1} \\
and \\
${}^\ddagger${\it
Faculty of Physics, University of Vienna\\
Boltzmanngasse 5, A-1090 Vienna, Austria}  }

\bigskip

\date{\today}

\textbf{Abstract}

\bigskip
\begin{minipage}{14cm}%

We consider a  solution of a IKKT-type matrix model 
which can be considered as a 1+1-dimensional  space-time with Minkowski signature and
a Big Bounce-like singularity.
A suitable $i\varepsilon$ regularization of the Lorentzian matrix integral is proposed, which leads to the standard $i\varepsilon$-prescription for the effective field theory. 
In particular, the Feynman propagator is recovered locally for late times.
This demonstrates that a causal structure and time evolution can emerge in
the matrix model, even on non-trivial geometries.
We also consider the propagation of modes  through  the Big Bounce,
and observe an interesting correlation between the post-BB and pre-BB
sheets, which reflects the structure of the brane in target space.

\end{minipage}

\end{center}

\end{titlepage}

\section*{Introduction}
\pdfbookmark[1]{Introduction}{Introduction}

Matrix theory can be viewed as an alternative approach to string theory.
There are two prominent matrix models which can be taken as starting point:
the BFSS model \cite{Banks:1996vh} is a model of matrix quantum mechanics
with a classical time variable, while the IKKT model \cite{Ishibashi:1996xs} is a pure matrix model without any
a priori notion of time. Both models admit solutions which can be interpreted in terms of
noncommutative D branes with a $B$ field, and fluctuations around such backgrounds lead to
noncommutative gauge theory.

The absence of a classical time variable in the IKKT model leads to an intriguing question: how can
time, and an effectively unitary time evolution, emerge from such a pure matrix model?
Indeed a naive interpretation of time in the noncommutative field theory leads to some issues,
which have been raised e.g. in \cite{Gomis:2000zz}.
However, to properly address this issue it is crucial to first identify the effective metric,
which is  dynamical in matrix models and depends on the background under consideration.
This can be clarified by studying the propagation of modes on such backgrounds  \cite{Steinacker:2010rh},
which allows to identify a unique effective metric closely related to the open string metric on the D-brane.
Only then a notion of time and time evolution can be identified.
Moreover, a proper treatment of the quantum theory can only be attempted in the maximally supersymmetric IKKT model.
From this perspective, the  objections raised in \cite{Gomis:2000zz} no longer apply.

In the present paper, we wish to elaborate some of these issues in more detail, and demonstrate that
a low-energy field theory can indeed emerge from IKKT-type matrix models
which displays the appropriate structures of causality and time evolution
required in quantum field theory. We will restrict ourselves to a free noncommutative scalar field theory
defined by a simplified model, i.e. ignoring loop corrections;
the latter should be addressed only in the full-fledged IKKT model.
More specifically, we will study a particular 1+1-dimensional solution of a reduced model, which can be viewed as a
toy model for the 3+1-dimensional covariant space-time solution given in \cite{Sperling:2019xar}.
The present solution is obtained as a projection of 2-dimensional fuzzy hyperboloid, with
structure reminiscent of  a 1+1-dimensional FLRW cosmology
with a Big Bounce (BB). It comprises a pre-BB and a post-BB
sector, which are glued together at the BB through a well-defined matrix 
configuration\footnote{In is interesting to note that a 1+1-dimensional space-time structure
linked by an Euclidean phase was recently reported in numerical studies of the bosonic 
IKKT model with mass term \cite{Nishimura:2022alt}.}.

The main claim of the present paper is that once a suitable definition of the matrix path integral in
Minkowski signature is implemented, the 2-point correlation functions have indeed the correct structure of a Feynman propagator
in quantum field theory. The Feynman $i\varepsilon$ structure is obtained from a suitable
regularization of the oscillatory matrix integral, which thus becomes absolutely  convergent and well-defined, at
least for finite-dimensional matrices. This prescription is slightly different from a similar regularization used
in recent computer simulations of the Lorentzian IKKT model
\cite{Nishimura:2019qal,Hatakeyama:2022ybs}, but is expected to be equivalent.

More explicitly, we obtain  the full set of (on- and off-shell) fluctuation
modes on the  FLRW-type  background under consideration. These modes stretch across the BB, and allow an explicit computation
of the Bogoliubov coefficients which relate the asymptotic pre- and post-BB regime.
Given these modes, we compute the propagator  by performing the matrix ``path'' integral, which displays the
standard structure of a Feynman propagator at times far from the BB.
This implies that the resulting effective field theory behaves as it should---at least at low energies---including
the appropriate causality structure and time evolution. 
In particular, the continuation of the modes across the BB suggests a
continuous time evolution across the mild singularity at the BB, with 
opposite ``arrow of time'' on the two sheets.
We also observe indications of some rather unexpected and intriguing correlations
between the pre-BB and  post-BB sheets.

The paper is organized as follows.  In section \ref{sec:definion-quantization}, we
define the matrix model and the $i\varepsilon$ prescription.  In section \ref{sec:fuzzy-hyperboloid}
we review the definition of a fuzzy 2-hyperboloid, explicitely construct harmonics on the
classical 2-hyperboloid and then use those to construct a harmonic basis for functions on
the fuzzy hyperboloid.  In section \ref{sec:Minkowski} we obtain our solution of interest,
a fuzzy two dimensional space with a Minkowski signature, $\cM^{1,1}$.  In section \ref{sec:fuzzy-scalar-field}
we describe dynamics of a single transverse fluctuation, solve the classical wave equation on
$\cM^{1,1}$ and study the Bogoliubov coefficients. Finally, in section
\ref{sec:path-integral} we put it all together, using the harmonic basis on the fuzzy hyperboloid
to compute a matrix model two point function in the background of an
emergent cosmological spacetime $\cM^{1,1}$.
Some further discussion is offered in section \ref{sec:discussion}.

\section{Definition of the model and quantization}
\label{sec:definion-quantization}

We will consider the following 3-dimensional IKKT--type matrix model 
\begin{align}
 S[Y] &= \frac 1{g^2}\Tr \Big(-[Y^a,Y^b][Y^{a'},Y^{b'}] \eta_{aa'} \eta_{bb'} \,
 - 2m^2 Y^a Y^b \eta_{ab}  \Big) \ . 
 \label{bosonic-action}
\end{align}
Here $\eta_{ab} = \diag(-1,1,1)$, and the $Y^a \in End(\cH)$ are hermitian  matrices
acting on some (finite- or infinite-dimensional) Hilbert space $\cH$.
Throughout this paper, indices will be raised and lowered with $\eta_{ab}$.
The action \eq{bosonic-action} is  a toy model for the
IKKT model \cite{Ishibashi:1996xs}, supplemented by a mass term $m^2$ which introduces a scale into the model
and without fermions  for simplicity.
This model has the  gauge invariance
\begin{align}
 Y^a \to U^{-1} Y^a U, \qquad U \in U(\cH) \ ,
\end{align}
which, as in Yang-Mills gauge theory, is essential to remove ghost contributions from the 
time-like direction, as well as a global $SO(2,1)$ symmetry.
The classical equations of motion are
\begin{align}
 \Box_Y Y^a = m^2 Y^a \ ,
 \label{eom-lorentzian-M}
\end{align}
where the matrix d'Alembertian
is defined as
\begin{align}
  \Box_Y = \eta_{ab} [Y^a,[Y^b,.]] \ .
  \label{Box-Y}
\end{align}
Equation \eq{eom-lorentzian-M} governs the propagation of scalar modes $\phi\in \End(\cH)$ on the 
background defined by  $Y^a$.
Such scalar modes arise 
in the matrix model from transverse fluctuations of the background solution, while
the tangential fluctuations give rise to gauge fields.
However, such gauge fields are not dynamical in 2 dimensions, and we will
focus on the scalar modes in the present paper.

Quantization of the model is defined via a matrix path integral,
\begin{align}
 Z = \int dY e^{i S[Y]}  \ .
 \label{path-integral}
\end{align}
As is the case with the oscillatory
path integral in Lorentzian QFT, this is not well defined as it stands. 
It was shown in 
\cite{Krauth:1998yu} that ,for pure bosonic 
Euclidean Yang-Mills matrix model, the matrix integral makes sense in $d\geq 3$ dimensions.
In the case of Minkowski signature, 
one possibility to define the path integral is to put an IR cutoff in both space-like
and time-like directions  as was done in \cite{Kim:2011cr}.
Here we propose a similar but more elegant regularization, giving the mass term
$\Tr( m^2 Y^a Y^b \eta_{ab})$  a suitable imaginary part
as follows:
\begin{align}
  \Tr (m^2 Y^a Y^b \eta_{ab} )
  &\to \Tr (  -(m^2 +i\varepsilon) Y^0 Y^0 + (m^2 -i\varepsilon) Y^i Y^i )   \ .
  \label{iepsilon-mass}
\end{align}
We thus define 
\begin{align}
 S_\varepsilon[Y] &= \frac 1{g^2}\Tr \Big([Y^0,Y^j]^2  \, 
    - [Y^i,Y^j]^2 + 2 (m^2 +i\varepsilon)(Y^0)^2 -2 (m^2 -i\varepsilon) (Y^j)^2\Big) .
 \label{MM-Mink}
\end{align}
which reduces to \eq{bosonic-action} in the limit $\varepsilon \searrow 0$.
Then, the integral 
\begin{align}
 Z_\varepsilon = \int dY e^{i S_\varepsilon[Y]}  
 \label{path-integral-eps}
\end{align}
is absolutely convergent
for any $\varepsilon > 0$.
To prove this, it suffices 
to observe  that
\begin{align}
 \int dY \left |e^{i S_\varepsilon[Y]} \right |  
 = \int dY \left |e^{\, \frac {2i}{g^2}\Tr\big((m^2 +i\varepsilon)(Y^0)^2 - (m^2 -i\varepsilon) (Y^j)^2\big)} \right | \ < \infty
 \label{path-integral-eps2}
\end{align}
since the rhs is a  Gaussian integral with good decay properties.
Note that the integration is always over the space of hermitian matrices $(Y^a)^\dagger = Y^a$, even
for the time-like matrices. 
In view of \eq{MM-Mink}, this regularization  amounts to Feynman's  $i\varepsilon$ -prescription in quantum field theory, and therefore  automatically 
imposes the appropriate causality structure in the propagators.
This will be verified explicitly in section \ref{sec:path-integral},
by computing the propagator in terms of the matrix path integral for a free scalar field.

\section{The fuzzy 2-hyperboloid \texorpdfstring{$H^2_n$}{H}}
\label{sec:fuzzy-hyperboloid}


In analogy to the well-known case of the fuzzy sphere $S^2_N$,
the fuzzy 2-hyperboloid $H^2_n$ \cite{Ho:2000br,Jurman:2013ota,Pinzul:2017wch}
is defined in terms of generators  
of $SO(1,2)$ acting on a unitary irreducible representation.
Let $\cM_{ab}$ be the generators of $\mso(1,2)$, which satisfy
\begin{align}
  [\cM_{ab},\cM_{cd}] &= i \left(\eta_{ac}\cM_{bd} - \eta_{ad}\cM_{bc} - 
\eta_{bc}\cM_{ad} + \eta_{bd}\cM_{ac}\right) \ .
 \label{M-M-relations-noncompact}
\end{align}
Fuzzy $H^2_n$ is then defined in terms of vector operators
$K_{a} := \frac 12\epsilon_{abc} \cM^{bc}$, which satisfy
\begin{align}
 [K_a,K_b] &= i\epsilon_{abc} K^c \nn  
\end{align}
using the convention $\epsilon_{012} = 1$.
Explicitly, $\cM_{12} = K_0$, $\cM_{20}= -K_1$ and $\cM_{01}= -K_2$
satisfy
\begin{align}
[K_1,K_2] &= -i K_0,\ \ [K_2,K_0] = i K_1,\ \ [K_0,K_1] = i K_2 \ .
\label{su11-algebra}
\end{align}
Here $K_0$ generates the compact $SO(2) \subset SO(1,2)$ subgroup, while
$K_1$ and $K_2$ generate non-compact $SO(1,1)\subset SO(1,2)$ subgroups.
As usual, it is convenient to introduce the ladder operators
\begin{align}
K_\pm = K_1\pm i K_2 \ ,
\label{ladderop}
\end{align}
which satisfy 
\begin{align}
\left[ K_0, K_\pm \right] =\pm K_\pm,\ \ \left[ K_+,K_-\right]=-2K_0.
\label{kpmalg}
\end{align}
The Casimir operator of $\mso(1,2)$ is defined as
\begin{align}
C^{(2)} = - \eta^{ab} K_a K_b =  - {K_1}^2 - {K_2}^2 + {K_0}^2 .
\end{align}

\subsection{Fuzzy \texorpdfstring{$H^2_n \subset \R^{1,2}$}{hyperboloid} as brane in target space}

For any unitary irrep $\cH$  of $SO(1,2)$,
define the hermitian generators
\begin{align}
 X^a &:= r K^{a}, \qquad a = 0,1,2  
\end{align} 
where $r$ is a parameter of dimension length. They satisfy 
\begin{align} 
 \eta_{ab} X^a X^b &=  - X^0 X^0 + X^1 X^1 +  X^2 X^2 = - r^2C^{(2)} \one \ ,\nn\\
   [X^a,X^b] &=  ir  \epsilon^{abc} X_c \ .
 \label{hyperboloid-constraint}
\end{align}
Moreover, it follows easily from these Lie algebra relations that
\begin{align}
 \Box_X X^a = [X_b,[X^b, X^a]] = -2 r^2 X^a, \qquad a=0,1,2 \ .
\end{align}
Therefore these $X^a$ provide a solution of the  matrix model \eq{bosonic-action} 
for\footnote{The negative mass does not imply an instability in the Minkowski case, it will 
merely lead to a cosmological solution with $k=-1$. The transverse fluctuations
will be stabilized by a positive mass in section \ref{sec:fuzzy-scalar-field}. }
\begin{align}
 m^2 = -2 r^2 \ .
 \label{mass-radius}
\end{align}
Finally we have to choose an appropriate representation. 
To obtain a one-sided hyperboloid, we should choose a discrete series 
positive-energy unitary irrep $\cH_n := D_n^+$ of $SO(2,1)$, 
as reviewed in appendix \ref{sec:unitary-reps}.
Then 
\begin{align} 
 -R^2 := \eta_{ab} X^a X^b &= X^i X^i - X^0 X^0  = - r^2 n(n-1) < 0
 \label{hyperboloid-constraint-2}
\end{align}
and $X^0 = r K^0 > 0$ has positive spectrum, given by
\begin{align}
 {\rm spec}(X^{0}) =  r\{n, n+1, ... \} \ . 
 \label{X0-discrete}
\end{align}
This structure will be denoted as $H^2_n$.

\paragraph{Semi-classical limit.}

The semi-classical limit of 
$H^2_n$ is obtained by replacing the generator $X^a$ with 
functions $x^a$ satisfying the constraint 
\begin{align}
 \eta_{ab} x^a x^b  = -(x^0)^2 + (x^1)^2 + (x^2)^2=  - R^2 < 0 \ ,
 \label{hyperboloid-constraint-sc}
\end{align}
and a $SO(2,1)$-invariant Poisson structure\footnote{Note that $r \sim R/n$ can be viewed as 
deformation parameter, which  for large $n$ separates the semi-classical regime from the noncommutative regime.}
\begin{align}
   \{x^a,x^b\} &=  r  \epsilon^{abc} x_c \ 
   \label{Poisson-bracket}
\end{align}
corresponding to \eq{hyperboloid-constraint}.
Accordingly, we can interpret the $X^a$ as quantized embedding functions of
a  one-sided Euclidean hyperboloid into $ \mso(2,1) \cong\R^{1,2}$,
\begin{align}
 X^a \sim x^a: \quad H^2 \hookrightarrow \R^{1,2} \ 
\end{align}
This is the quantization of the coadjoint 
 orbit $H^2$ of $SO(2,1)$, with the $SO(1,2)$- invariant 
 Poisson bracket (or symplectic structure) \eq{Poisson-bracket}.
The operator algebra $\End(\cH_n)$ can thus be interpreted as quantized 
algebra of functions on $\cC^\infty(H^2)$. 
Clearly $H^2_n$ has a finite density of microstates, according to the 
Bohr-Sommerfeld rule.

\subsection{Functions and harmonics on classical \texorpdfstring{$H^2$}{hyperboloid}}
\label{sec:harmonics-H2}

The action of $SO(2,1)$ on functions $\phi \in \cC^\infty(H^2)$ is realized via the
Hamiltonian vector fields
\begin{align}
 K^a \triangleright\phi =  \frac ir\{x^a,\phi\} \ 
 = -\frac 12\epsilon^{abc} \cM_{bc}\triangleright\phi   \ .
\end{align}
In particular, the space of square-integrable functions $\phi(x)$ on $H^2$ forms a unitary representation, which 
decomposes into unitary irreps of $SO(2,1)$.
It follows that the  Casimir 
\begin{align}
 C^{(2)} \phi &= - K^a K_a \phi  = (K_0^2 - K_1^2 - K_2^2 ) \phi 
  =: - \frac 1{r^2} \Box_H \phi \nn\\
  \Box_H \phi &:= - \{x_a,\{x^a,\phi\}\} 
\label{Box-H-def}
\end{align}
coincides with the metric Laplacian $\Delta_H$ on $H^2$ up to a factor, 
\begin{align}
 \Delta_H \phi = \frac 1{R^2} C^{(2)} \phi
 =  - \frac 1{R^2 r^2} \Box_H \phi
 &= \frac{1}{\sqrt{|g_{\mu\nu}|}}\del_\mu\left(\sqrt{|g_{\mu\nu}|}\, g^{\mu\nu}\del_\nu \phi\right) \ ,
 \label{Casimir-Laplace}
\end{align}
where $g$ is the induced metric on $H^2$.
This  gives  $C^{(2)} P_l(x) = l(l+1) R^2 P_l(x)$ for irreducible polynomials of degree $l$ in $x^a$; 
for example,
\begin{align}
  C^{(2)} x^b =   \frac 1{r^2} \{x_a,\{x^a,x^b\}\} =  2 x^b \ .
\end{align}
For square-integrable functions, the Casimir must  be negative definite,
which is indeed the case for functions in the principal series irreps.

\paragraph{Hyperbolic coordinates and eigenfunctions.}

To find the general eigenfunctions of $\Delta_H$, consider the following coordinates\footnote{These coordinates
are compatible with the projection to $\cM^{1,1}$ considered below.} 
on $H^2$:
\begin{align}
\begin{pmatrix}
 x^0 \\ x^1 \\ x^2 
\end{pmatrix}
 = R\, \begin{pmatrix}
  \cosh(\eta)\begin{pmatrix}
  \cosh(\chi) \\
   \sinh(\chi) \end{pmatrix}  \\
   \sinh(\eta)
\end{pmatrix}              \   ,                             
 \label{local-hyperbolic-coords}
\end{align}
for $\eta,\chi\in\R$.
Then, the induced metric on $H^2$ is 
\begin{align}
 ds^2|_H &= g_{\mu\nu} dx^\mu dx^\nu = R^2(d\eta^2 + \cosh^2(\eta)d\chi^2) 
 \label{ind-metric-H2}
\end{align}
with $\sqrt{g} = R^2\cosh(\eta)$.
Hence the metric Laplacian on $H^2$ is given by
\begin{align}
  \Delta_H \phi &= \frac{1}{\sqrt{|g_{\mu\nu}|}}\del_\mu\big(\sqrt{|g_{\mu\nu}|}\, g^{\mu\nu}\del_\nu \phi\big)
  = \frac{1}{R^2}\left(\del_\eta^2\phi + \tanh(\eta)\del_\eta\phi
  +  \frac{1}{\cosh^2(\eta)}\del^2_\chi \phi\right) \ .
  \label{Laplace-diffop}
\end{align}
Now consider eigenfunctions of $\Delta_H$:
\begin{align}
 \Delta_H \phi = \l \phi \ .
 \label{Laplace-H2-equation}
\end{align}
The separation ansatz 
\begin{align}
 \phi(\eta,\chi) = f(\eta) e^{i k \chi}
\end{align}
leads to  
\begin{align}
  \big(\del_\eta^2 + \tanh(\eta)\del_\eta -  \frac{k^2}{\cosh^2(\eta)} - R^2 \l\big) f  = 0 \ .
  \label{f-eta-eq}
\end{align}
To bring this to standard form, we
can substitute $u=\tanh \eta \in (-1,1)$ 
and define  $f(u) = (1-u^2)^{1/4} h(u)$, to obtain
\begin{align}
  (1 - u^2) h'' - 2 u h' + \left (-\left(k^2 + \tfrac 14\right) - \frac{\frac 14 +  \l R^2}{1-u^2} \right) h   &=  0 \ . 
\label{legendre-eq}
\end{align}
The solutions are associated Legendre functions of the
first and second kind, 
$\mathsf{P}_\nu^{\mu}$ and  $\mathsf{Q}_\nu^{\mu}$,
with
 \begin{align} 
 \qquad \nu(\nu+1) = - k^2- \frac 14 ~~~\mathrm{and} \qquad \mu^2 = \frac 14 + \l R^2 ~ .
 \label{Legendre-parameters-1}
\end{align}
We  use the definitions and conventions given in \cite{NIST:DLMF},
and all properties of these functions we require can be found therein.
The first relation amounts to\footnote{Strictly
speaking, it should be $\nu = -\half  \pm ik$, but
as we will use associated Legendre functions of the 
first kind as our basis, this 
is irrelevant since $\mathsf{P}^\mu_\nu = \mathsf{P}^\mu_{-1-\nu}$.}
\begin{align}
  \nu(k) = -\frac 12 +  i |k| \ .
   \label{degree}
\end{align}
For $\l < -\frac{1}{4 R^2}$,  the solutions  
realize the principal series irreps $P_s$ 
with
\begin{align}
s= |\mu| = \sqrt{ - \l R^2 - \quarter}~~~\mathrm{is~positive}.
 \label{order:principal}
 \end{align}
Indeed, the Casimir is 
\begin{align}
 C^{(2)} = R^2 \l  = -(s^2 + \frac 14) < - \frac 14 
\end{align}
using \eq{Casimir-Laplace}, which corresponds precisely to the principal series \eq{principal}.

For $\l \in (-\frac{1}{4R^4},0)$, 
the solutions  correspond to the complementary series irreps $P_j^c$ 
with $j=\half + \sqrt{(\quarter + \l R^2)}$ from equation \eq{complementary}, since
 $C^{(2)} = R\lambda^2 = j(j-1) \in (-\frac 14,0)$.

\paragraph{Principal series solutions and asymptotics.}

For $\mu^2 < 0$, the differential
equation (\ref{legendre-eq}) has two linearly independent
solutions, corresponding to the principal series.
It will be  convenient to use $\mu = \pm i s$, so that these solutions
are\footnote{$\mathsf{Q}^{\pm is}_{\nu(k)}$ can be written
  as linear combinations of $\LP^{\pm is}_{\nu(k)}$ and is therefore
  not an independent solution.
We can use either $\mathsf{P}^{is}_{\nu}(-u)$ or $\mathsf{P}^{is}_{\nu}(u)$, since the 
equation is invariant under $u \to -u$.}
$\mathsf{P}^{is}_{\nu(k)}(-u)$ and $\mathsf{P}^{-is}_{\nu(k)}(-u)$
for every positive $s$.
For later use, we consider their asymptotic behavior.  
As $x\rightarrow 1^-$, we have 
\begin{align}
\mathsf{P}^{\mu}_{\nu(k)}(x) ~~
\stackrel{x \to 1^-}{\sim}~~
\frac{1}{\Gamma\left(1 - \mu \right)}
\left(\frac{2}{1-x}\right)^{\mu/2} ~ .
\label{asymptotics-1}
\end{align}
Therefore
\begin{align}
\mathsf{P}^{\pm is}_{\nu(k)}\left(-u\right) \stackrel{u \to -1^+}{\sim}
\frac{1}{\Gamma\left(1 \mp is\right)}
\left(\frac{2}{1+u}\right)^{\pm is/2} ~
\end{align}
for  $u \rightarrow -1^+$,
or equivalently
\begin{align}
\mathsf{P}^{\pm is}_{\nu(k)}\left(-\tanh \eta \right) 
~\stackrel{\eta \to-\infty}{\sim}~
\frac{ e^{ \mp is \eta}}{\Gamma\left(1 \mp is\right)}~ .
 \label{P-asympt-neg}
\end{align}
Hence these solutions behave like plane waves for $\eta \to -\infty$.

To obtain the behaviour of the solutions for $\eta \to \infty$,
we  use the following identity:
\begin{align}
\frac{\sin\left((\nu-\mu)\pi\right)}{\Gamma\left(\nu+\mu+1\right)}~\mathsf{P}^{\mu}_{\nu}
\left(x\right)=\frac{\sin\left(\nu\pi\right)}{\Gamma\left(\nu-\mu+1\right)}
~\mathsf{P}^{-\mu}_{\nu}\left(x\right)-\frac{\sin\left(\mu\pi\right)}
{\Gamma\left(\nu-\mu+1\right)}~\mathsf{P}^{-\mu}_{\nu}\left(-x\right)~.
\end{align}
We can thus write 
\begin{align}
  \mathsf{P}^{-\mu}_{\nu}\left(-x\right) ~~=~~
&\frac{\sin\left(\nu\pi\right)}{\sin\left(\mu\pi\right)}~\mathsf{P}^{-\mu}_{\nu}
\left(x\right)~-~\frac{\sin\left((\nu-\mu)\pi\right)}{\sin\left(\mu\pi\right)}
\frac     {\Gamma\left(\nu-\mu+1\right)}{\Gamma\left(\nu+\mu+1\right)}
~\mathsf{P}^{\mu}_{\nu}\left(x\right)
\end{align}
and asymptotically
\begin{align}
  \mathsf{P}^{-\mu}_{\nu}\left(-x\right) ~~~
\stackrel{x \to 1^-}{\sim}~~~
&\frac{\sin\left(\nu\pi\right)}{\sin\left(\mu\pi\right)}~
\frac{1}{\Gamma\left(1 +\mu \right)}
\left(\frac{2}{1-x}\right)^{-\mu/2}
\nn\\
&-~~~\frac{\sin\left((\nu-\mu)\pi\right)}{\sin\left(\mu\pi\right)}
\frac     {\Gamma\left(\nu-\mu+1\right)}{\Gamma\left(\nu+\mu+1\right)}
\frac{1}{\Gamma\left(1 -\mu \right)}
\left(\frac{2}{1-x}\right)^{\mu/2} \ .
\label{P-asympt}
\end{align}
Therefore, for $\eta \to \infty$, we have 
\begin{align}
\mathsf{P}^{-\mu}_{\nu(k)}\left(-\tanh \eta \right) 
\stackrel{\eta \to \infty}{\sim}~~~
&\frac{\sin\left(\nu\pi\right)}{\sin\left(\mu\pi\right)}~
\frac{1}{\Gamma\left(1 +\mu \right)}
e^{-\mu\eta}~~-
\nn\\
&\frac{\sin\left((\nu-\mu)\pi\right)}{\sin\left(\mu\pi\right)}
\frac     {\Gamma\left(\nu-\mu+1\right)}{\Gamma\left(\nu+\mu+1\right)}
\frac{1}{\Gamma\left(1 -\mu \right)}
e^{\mu\eta} \nn \\
 = ~~-~~&\frac{\sin\left(\nu\pi\right)\Gamma\left(-\mu \right)}{\pi}~
~e^{-\mu\eta}~~+~~
\frac     {\Gamma(\mu)}{\Gamma\left(\mu-\nu\right)\Gamma\left(\nu+\mu+1\right)}
~~e^{\mu\eta}
 \label{P-asympt-pos}
\end{align}
using $\Gamma(z)\Gamma(1-z)\sin(\pi z) = \pi$.

To summarize, a complete set of solutions of \eq{Laplace-H2-equation} is given by 
\begin{align}
\boxed{ \ \ 
 \Upsilon^{s\pm}_k(\eta,\chi) := \frac 1{\sqrt{\cosh\eta}}\,
 e^{i k \chi} \LP^{\pm is}_{-\frac 12 + i |k|}(-\tanh(\eta)) 
 \quad \mbox{for} \quad s > 0, \ k\in\R
 \ \ } \  .
 \label{Upsilon}
\end{align}
These $\Upsilon^{s\pm}_k$  realize the principal series irrep $P_s$ \eq{principal}.
They are the analogs of the spherical harmonics, and 
the space  of all square-integrable functions on $H^2$ is spanned by
the $\Upsilon^s_k$. 
We will find analogous solutions 
in the Minkowski case (see Section \ref{sec:Minkowski})
corresponding to propagating  waves, where
$\mathsf{P}^{\pm is}_{\nu(k)}(-u)$ will be interpreted as 
positive ($\mathsf{P}^{is}$) and negative ($\mathsf{P}^{-is}$)
frequency  modes in the far past.

\paragraph{Comment on the complementary series.}

We have seen that $s^2 > 0$ (or equivalently $\mu^2<0$) is  the case where the functions oscillate 
for $\eta \to \pm \infty$.
In contrast, the solutions with $s^2 < 0$  corresponding to 
the complementary series do not describe  waves propagating 
in the far past or future. For this reason, we will not consider the 
complementary series solutions any further.

\subsubsection{Symplectic form, integration and inner product}

The $SO(2,1)$-invariant volume form (i.e. the symplectic form) 
is given by 
\begin{align}
\omega = \frac{R}{r} \cosh(\eta) d\eta d\chi \ ,
\label{symplectic-form}
\end{align}
corresponding to the Poisson bracket
\begin{align}
  \{\eta,\chi\} = -\frac{r}{R\cosh(\eta)} \ .
  \label{eta-chi-bracket}
\end{align}
This is consistent with $\sqrt{|g|} = R\cosh(\eta)$ in the $\eta\chi$
coordinates, \eq{ind-metric-H2}.
The trace corresponds to the integral over the symplectic volume form on $H^2$,
\begin{align}
 2\pi \: \Tr(\hat \phi) \sim \int_{H^2} \omega \phi(x)  \ .
\end{align}
In particular, we can define an  $SO(2,1)$-invariant inner product via 
\begin{align}
 \langle \phi, \psi\rangle &:=\int_{H^2} \omega\,\phi^* \psi \ ,
 \label{inner-symp}
\end{align}
which  defines the space $L^2(H^2)$ of
square-integrable functions.
Then the eigenmodes \eq{Upsilon} of $\Box$ satisfy orthogonality relations
\begin{align}
\langle \Upsilon^{s'\pm}_{k'}, \Upsilon^{s\pm}_k\rangle &:=
 \int_{H^2} \omega\, (\Upsilon^{s'\pm}_{k'}(\eta,\chi))^* \Upsilon^{s\pm}_k(\eta,\chi) \nn\\
  &= \frac{R}{r}\int d\eta d\chi
  e^{-i k' \chi} (\LP^{\pm is'}_{-\frac 12 + i |k'|}(-\tanh(\eta)))^*
  e^{i k \chi} \LP^{\pm is}_{-\frac 12 + i |k|}(-\tanh(\eta)) \nn\\
  &=\frac{R}{r}\int  d\chi e^{i (k-k') \chi} \int du \frac{1}{1-u^2}
   (\LP^{\pm is'}_{-\frac 12 + i |k'|}(-u))^*
   \LP^{\pm is}_{-\frac 12 + i |k|}(-u) 
  \label{orth-H2-explicit}
\end{align}
using $d\eta = \frac{du}{1-u^2}$. 
The last integral  can be evaluated explicitly 
using the orthogonality relations \eq{Ipm-orth}
if desired.

\subsection{Functions on fuzzy \texorpdfstring{$H^2_n$}{hyperboloid} and coherent states}
\label{sec:coherent-Q}

\paragraph{Tensor product decomposition}

The fuzzy analog of the algebra of functions  $\cC^\infty(H^2)$ is given by
$\End(\cH_n)$. To understand the fluctuation spectrum, we should decompose this into irreps 
of $SO(2,1)$. This is somewhat non-trivial since these are infinite-dimensional representations,
as in the commutative case. However, we can use the fact that
$SO(2,1)$ acts on noncommutative functions $\hat\phi$ via the adjoint
\begin{align}
 K^a \triangleright\hat\phi = [K^a,\hat\phi] =  \frac 1r [X^a,\hat\phi] \ , \qquad \hat\phi \in \End(\cH_n) \ .
\end{align}
Square-integrable functions $\phi \in L^2(H^2)$
correspond to Hilbert-Schmidt operators 
$\hat\phi \in \End(\cH_n)$, which form a Hilbert space, and accordingly
decompose into unitary irreps of $SO(2,1)$, defining
fuzzy scalar harmonics $ \hat \Upsilon^{s\pm}_k$.

The decomposition of Hilbert-Schmidt operators in $\End(\cH_n)$
 is obtained from the unitary tensor product decomposition \cite{repka1978tensor}:
\begin{align}
 \End(\cH_n) \cong
 D_n^+ \otimes D_n^- \cong \int_0^\infty ds P_s \ .
 \label{decompo}
\end{align}
The $P_s$ are principal series irreps which asymptotically correspond to
plane waves, and the direct integral on the rhs means that
square-integrable 
functions are obtained as usual by forming wave-packets of these.

\paragraph{Coherent states and an isometric quantization map.}

Due to the above unique decomposition, the quantization map between $\cC^\infty(H^2)$ to $\End(\cH_n)$ is
fixed by symmetry up to a set of normalization constants.  To make this more
explicit, will can use coherent states.  These are defined in a natural way using the
fact that $\cH_n$ is a  lowest weight representations.
Let 
\begin{align}
 |x_0\rangle := |n,n\rangle \qquad \in \cH_n
\end{align}
be the (unit length)
lowest weight state. This is an optimally localized state at the  ``south pole''  $x_0 = (R,0,0)\in\R^{1,2}$ of $H^2$.
Then the  coherent state 
\begin{align}
 |x\rangle:= U_g|x_0\rangle\qquad \in \cH_n
\end{align}
is defined by acting with a $SO(2,1)$ rotation $U_g$
which rotates $x_0$ into $x\in H^2$. The ambiguity in the choice of the  group element
$g\in SO(2,1)$ leads to a $U(1)$ phase ambiguity, so that the coherent states form a $U(1)$ bundle over $H^2$.

With this, we can define any $SO(2,1)$-equivariant
quantization map $\cQ$ through its action on the harmonics
\begin{align}
 \cQ: \quad L^2(H^2) &\to \End(\cH_n)  \nn\\
  \Upsilon^{s\pm}_k &\mapsto \hat\Upsilon^{s\pm}_k
  :=   c_s  \int\limits_{H^2}  \omega \, \Upsilon^{s\pm}_k(x) 
  \left|x\right\rangle \! \left\langle x\right| \ ,
  \label{quantization-isometry}
\end{align}
where $c_s$ are (so far) undefined constants.    

This map is  one-to-one as a map from
square-integrable functions to Hilbert-Schmidt operators, and
its inverse is given  by the symbol
\begin{align}
 \hat \Upsilon^{s\pm}_k \mapsto  \Upsilon^{s\pm}_k(x) =  d_s
\langle x | \hat \Upsilon^{s\pm}_k  | x\rangle \ .
 \label{quantization-inverse}
\end{align}
where the coefficients $d_s$ satisfy
\begin{align}
\label{Q-inverse-coefficients}
  c_s d_s   \int\limits_{H^2} \omega \,  \Upsilon^{s\pm}_k(x) \, |\langle x | x' \rangle |^2
& = \Upsilon^{s\pm}_k(x') \ . \end{align}
Since $\cQ$ respects $SO(2,1)$, it is an intertwiner of its generators
\begin{align}
  [X^a,\cQ(\phi)] &= \cQ(i\{x^a,\phi(x)\}) \ ,
  \label{Q-intertwiner}
\end{align}
so that
the Laplacian is respected as well:
\begin{align}
 \Box_X \cQ(\phi) = [X^a,[X_a,\phi]]  = \cQ(\Box_H \phi) \ .
 \label{Laplacian-intertwiner}
\end{align}
Here $\Box_H$  \eq{Casimir-Laplace}
is the usual Laplacian on $H^2$,
which is essentially the quadratic Casimir.

When the coefficients $c_s$ are all equal, this construction is
the well known quantization map used, for example, on symmetric spaces,
\begin{align}
 \quad \cC(H^2) &\to \End(\cH_n)  \nn\\
 \phi(x)  &\mapsto \hat \phi = \int\limits_{H^2} \omega \, \phi(x) 
\left|x\right\rangle \!  \left\langle x\right| \ .
 \label{quantization-map}
\end{align}

Here, however, we are interested in a quantization which is
an isometry with respect to the inner products defined by the 
trace and \eq{inner-symp}, respectively.  This can be accomplished
chosing suitable normalization constant $c_s$ for each $\hat \Upsilon^{s\pm}_k$, such that
\begin{align}
  \langle \Upsilon^{s\pm}_k, \Upsilon^{s\pm}_{k'}\rangle & =\int\limits_{H^2} \omega\,
  (\Upsilon^{s\pm}_k)^* \Upsilon^{s\pm}_{k'} = 2\pi \: \Tr \left (
  (\hat \Upsilon^{s\pm}_k)^\dagger \hat \Upsilon^{s\pm}_{k'} \right )
  \label{isometry-condition}
\end{align}

When $\cQ$ is an isometric map, we must have $ d_s = 2\pi c_s$.  Coefficients $c_s$ can
therefore be computed from equation \eq{Q-inverse-coefficients}.

\section{1+1-dimensional (squashed) space-time \texorpdfstring{$\cM^{1,1}$}{M1,1}}
\label{sec:Minkowski}

\begin{figure}[h]
\hspace{2cm}
 \includegraphics[width=0.5\textwidth]{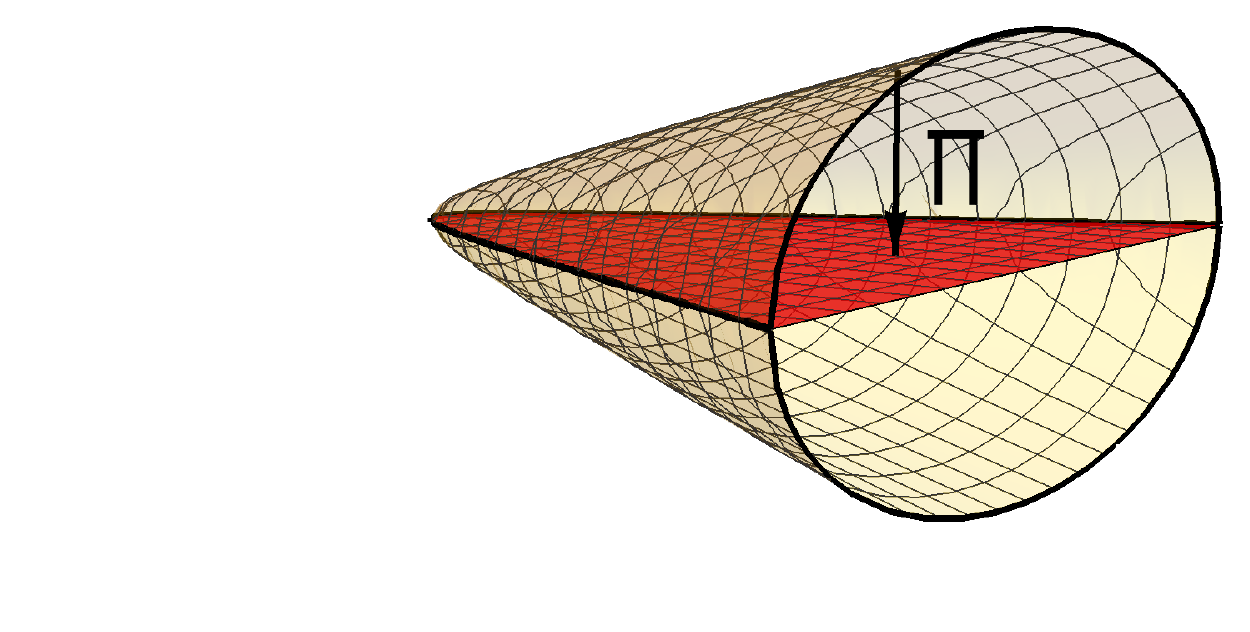}
 \caption{Sketch of the projection $\Pi_x$ from $H^2$ to $\cM^{1,1}$  with 
Minkowski signature.}
 \label{fig:projection}
\end{figure}

Following \cite{Steinacker:2017bhb}, we can obtain 
a space with Minkowski signature by
projecting of $H^2$ onto the $0,1$ plane as follows
\begin{align}
 \Pi: \quad H^{2} \ &\to \R^{1,1}  \subset  \R^{1,2}, \nn\\
               (x^0,x^1,x^2) &\mapsto (x^0,x^1,0) .
 \label{projection}
\end{align}
The projected space  $\cM^{1,1} = \cM^+ \cup \cM^-$ consists of two sheets which are connected at the boundary,
cf. figure \ref{fig:projection}.
This respects the $SO(1,1)$ generated by $K_2$.
In the fuzzy case, this projection is realized simply by dropping $X^2$ from the matrix background, 
and considering a new background through $X^0$ and $X^1$ only.
Thus define\footnote{Note that $X^0$ and $X^1$ generate the full algebra of functions 
$\End(\cH)$ on $H^2_n$; only  the effective geometry defined by the matrix background is changed.}
\begin{align}
   Y^\mu &=  X^\mu , \quad \mbox{for} \ \  \mu = 0,1 \ .
 \label{H4-ansatz}
\end{align}
 The $\mso(2,1)$ algebra  gives
\begin{align}
  [Y^0,[Y^0,Y^1]] &= i r^3 [K^0, K^2] = r^2 Y^1  \nn\\
  [Y^1,[Y^1,Y^0]] &= -i r^3 [K^1, K^2] = - r^2 Y^0  
\end{align}
so that\footnote{Note that $Y^\mu$ must be eigenvectors of $\Box_Y$ 
due to $SO(1,1)$ invariance.}
\begin{align}
 \Box_Y Y^\mu = -[Y^0,[Y^0,Y^\mu]]+[Y^1,[Y^1,Y^\mu]] &= - r^2 Y^\mu \ , \qquad \mu = 0,1 .
 \label{Box-Y-M11}
\end{align}
 This means that the $Y^\mu$ for $\mu = 0,1$  provide a solution of the Lorentzian matrix model \eq{eom-lorentzian-M}
with positive mass
\begin{align}
m^2 =   r^2 . 
\label{radius-mass-proj}
\end{align}
This is the solution of interest here, which can be realized either in a 1+1-dimensional matrix model, or in the 
3 (or higher)- dimensional model \eq{bosonic-action} by setting
the remaining $Y^a$ to zero.
If we keep such extra matrices in the model, their fluctuations 
will play the role of scalar fields on the background, viewed as transverse fluctuations of the brane. 
This will be discussed in section \ref{sec:fuzzy-scalar-field}.
Note that $m^2>0$ suggests stability of this background, which  should be 
studied in more detail  elsewhere.

$Y^\mu$ transform as vectors of $SO(1,1)$, which can be 
realized by the adjoint i.e. through gauge transformations.
Hence the solution admits a global $SO(1,1)$ symmetry. 
In the semi-classical limit, this defines a foliation of $\cM$ into 
one-dimensional space-like hyperboloids $H^1_t$, more precisely
\begin{align}
\cM ^{1,1}\ = \ \cM^+ \cup \cM^- \ =  H^1_{t_0} \ \cup \  2\bigcup\limits_{t > t_0} H^1_{t} \nn
\end{align}
one for each sheet except for $t=t_0 = R$.
The two sheets $\cM^+ \cup \cM^-$ are connected at $t=R$, 
cf. figure \ref{fig:projection}. We will see that
the $x^0$ direction is time-like, and that $\cM^{1,1}$ 
resembles a double-covered 1+1-dimensional
FLRW space-time with hyperbolic ($k=-1$) spatial geometry, similar to that in \cite{Steinacker:2017bhb}.
Note that these time-slices are infinite in the space direction, 
even at the Big Bounce $t=t_0$. 
Therefore it is not unreasonable to expect a unitary 
time-evolution for all $t$.

\subsection{Semi-classical geometry}

\paragraph{Induced metric.}

Consider the semi-classical limit $Y^\mu \sim y^\mu$.
On this projected space, 
the induced metric on $\cM^{1,1}\subset \R^{1,1}$ is clearly Lorentzian,
\begin{align}
 g_{\mu\nu} = (-1,1) = \eta_{\mu\nu} , \qquad \mu,\nu=0,1 \ 
\end{align}
in Cartesian coordinates $y^\mu$.
This is recognized as a $SO(1,1)$-invariant FLRW metric with $k=-1$,
by decomposing $\cM$ into 1-hyperboloids $H_t$
\begin{align}
\begin{pmatrix}
 y^0 \\ y^1 
\end{pmatrix}
 = t\, \begin{pmatrix}
  \cosh(\chi) \\
   \sinh(\chi)
\end{pmatrix}                           
 \label{local-hyperbolic-coords-2}
\end{align}
for $t = R \cosh(\eta)\in [R,\infty)$. In particular,
\begin{align}
 t^2 = - y^\mu y^\nu \eta_{\mu\nu} \geq R^2, \qquad 
 R^2 = t^2 - x_2^2 > 0
 \label{time-def}
\end{align}
where 
\begin{align}
x_2 = R\sinh(\eta) = \pm\sqrt{t^2-R^2} 
\end{align}
is a function on $H^2$ which allows to distinguish the two sheets of $\cM^{1,1}$ for $\eta\in\R$.
This gives the 2D flat Milne metric:
\begin{align}
 ds_g^2 &= -d t^2 + t^2\, d\chi^2 = -dy_0^2 + dy_1^2 \ .
 \label{milne-metric}
\end{align}
Here $\chi\in(-\infty,\infty)$ parametrizes the $SO(1,1)$-invariant space-like $H^1$ with $k=-1$.
The $(\eta, \chi)\in \R^2$ variables are very useful because they parametrize
both sheets of the projected hyperboloid $H^2$.

The induced metric $g$ can be viewed as closed-string metric in target space.
However as familiar from matrix models \cite{Steinacker:2010rh} and string theory \cite{Seiberg:1999vs}, 
the fluctuation on the brane are governed by a different metric or kinetic term:

\paragraph{Effective generalized d'Alembertian.}

We will see in the next section that
the kinetic term for a (transverse) scalar field on this background in the matrix model is governed by
\begin{align}
  \Box_y = -\{y^\mu,\{y_\mu,.\}\} = \Box_H + \{y_2,\{y_2,.\}\}
  \label{Box-y}
\end{align}
where $\Box_H$ is the Laplacian \eq{Box-H-def} on $H^2$.
The extra term is evaluated  easily as 
\begin{align}
 \{y_2,\{y_2,\phi\}\} 
  &= r^2\del^2_\chi\phi
\end{align}
using \eq{eta-chi-bracket}. Together with  \eq{Laplace-diffop} we obtain
\begin{align}
 \Box_y  \phi
 &=  -r^2 \Big(\del_\eta^2 + \tanh(\eta)\del_\eta
   - \tanh^2(\eta)\del^2_\chi  \Big) \phi \nn\\
    &= -\Big(\g^{\mu\nu}\del_\mu\del_\nu + O(\del)\Big)\phi \ .  
\label{BoxY-tchi}
 \end{align}
This is a  second-order hyperbolic differential operator with leading symbol
 $\g^{\mu\nu}p_\mu p_\nu$ where
 \begin{align}
  \g^{\mu\nu} = r^2\begin{pmatrix}
                 1 & 0 \\
                 0 & -\tanh^2(\eta)
                \end{pmatrix}
 \end{align}
in $(\eta\chi)$ coordinates.
This governs the propagation
of scalar fields on $\cM^{1,1}$, and respects the $SO(2,1)$ symmetry of a $k=-1$ FLRW space-time with time  $\eta$.
We also note the identity
\begin{align}
 \omega =  \rho\,\sqrt{|\g_{\mu\nu}|}\, d\xi^0 d\xi^1 \quad \mbox{with} \quad \rho = r R |\sinh(\eta)| \
 \label{omega-dilaton-id}
\end{align}
in local coordinates $\xi^\mu$.
In dimensions larger than 2, such a ``matrix Laplacian'' can always be written in terms of a metric
Laplacian (or d'Alembertian)
for a unique effective metric \cite{Steinacker:2010rh}. This is not possible in 2 dimensions
due to Weyl invariance\footnote{The conformal factor could be determined  for
gauge fields, which arise from tangential fluctuations of $\cM^{1,1}$ in the matrix model.
However we refrain from pursuing this direction in the present paper.}.
We will therefore study the  operator $\Box_y$ directly, which
 will be referred to as generalized d'Alembertian.
The metric $-\g^{\mu\nu}$ is that of a FLRW space-time
and clearly governs the local propagation and causality structure,
which is the main focus of the present paper.  
However it should not be considered as  effective metric.
The origin of $\g^{\mu\nu}$  will become clear in the next section.

\section{Scalar harmonics on fuzzy \texorpdfstring{$\cM^{1,1}$}{M1,1}}
\label{sec:fuzzy-scalar-field}

\subsection{Transverse fluctuations in the matrix model}

Scalar fields on $\cM^{1,1}$ are realized by the transverse (space-like) matrix
$Y^a, \ a=2$
in the model \eq{bosonic-action} or \eq{MM-Mink} (possibly extended by further matrices $Y^a$):
\begin{align}
 S[Y] &= S[Y^\mu]  + 
 \frac 2{g^2}\Tr \Big( -[Y^\mu,Y^a][Y_\mu,Y^a] - (m_\phi^2 - i \varepsilon) Y^a Y_a\Big) 
 \label{MM-Mink-extra}
\end{align} 
Here we include an arbitrary scalar mass parameter $m^2_\phi$, 
independent of $m^2$ in \eq{bosonic-action}.
We focus on one such transverse  matrix $Y^a =: \hat\phi$, viewed as scalar field on  $\cM^{1,1}$.
Its effective action is accordingly
\begin{align}
 S[\hat\phi] &= \frac 2{g^2}\Tr \Big( \hat\phi\Box_Y\hat\phi - (m_\phi^2 - i \varepsilon) \hat\phi^2\Big)
   = \frac 2{g^2} \int\limits_\cM \omega \Big( \phi \Box_y\phi - (m_\phi^2 - i \varepsilon) \phi^2\Big) =: S[\phi]
 \label{MM-Mink-scalar}
\end{align}
with $\Box_Y$  for matrices given in \eq{Box-Y-M11}, and $\Box_y$ 
for functions in \eq{Box-y}.
Here the matrix model $S[\hat\phi]$ is identified with the semi-classical action $S[\phi]$ , which needs some explanation. The matrix $\hat\phi\in\End(\cH_n)$ is identified via the quantization map
$\cQ$ \eq{quantization-isometry}
with a function $\phi\in L^2(H^2)$, which in turn is identified with $L^2(\cM^{1,1})$
via  $\Pi$ \eq{projection}. The symplectic form $\omega$ is also the same as on $H^2$, so that 
the integral over $\cM^{1,1}$ can be viewed as an integral over $H^2$.
As discussed in section \ref{sec:coherent-Q}, $\cQ$ is (by definition) an isometry between 
$L^2(H^2) = L^2(\cM^{1,1})$    
and (Hilbert-Schmidt operators in) $\End(\cH_n)$.
Moreover, the $[Y^\mu,.]$ are $SO(2,1)$ generators which commute with the quantization map 
$\cQ$. Therefore the free matrix model $S[\hat\phi]$ is identically mapped by $\cQ$ 
to the classical action $S[\phi]$.

In the same vein, the matrix equation of motion for the scalar field 
\begin{align}
 \Box_Y \hat\phi =  m^2_\phi \hat\phi
\end{align} 
is  equivalent to the semi-classical (Poisson) wave equation 
\begin{align}
 \Box_y \phi =  m^2_\phi \phi  \ .
 \label{eom-phi-class}
 \end{align}
We will determine the classical eigenmodes of $\Box_y$ explicitly below.

To understand the role of $\g^{\mu\nu}$ in \eq{BoxY-tchi}, it is instructive to rewrite the above kinetic term
as follows
\begin{align}
  S[\phi] &=\frac 2{g^2} \int\limits_\cM \omega \Big( \g^{\mu\nu} \del_\mu\phi\del_\nu\phi - (m_\phi^2 - i \varepsilon) \phi^2\Big)
\end{align}
in terms of a frame   \cite{Steinacker:2020xph}
\begin{align}
 E^{a\mu} = \{y^a,\xi^\mu\} , \qquad  \g^{\mu\nu} = \eta_{ab}E^{a\mu} E^{b\nu}
\end{align}
in any local coordinates $\xi^\mu$. In view of \eq{omega-dilaton-id}, this can be interpreted as action for a scalar field non-minimally coupled to a dilaton \cite{Callan:1992rs},
and it explains the origin and the significance of the
metric $\g^{\mu\nu}$. In the case of $3+1$ dimensions, this metric  turns out to be conformally equivalent to the effective metric \cite{Battista:2022hqn}.

\subsection{Eigenfunctions of \texorpdfstring{$\Box_y$}{box-y}}

We  want to solve the eigenvalue equation 
\begin{align}
  \Box_y \phi = \l \phi \ , 
  \label{Box-eigenfunctions}
\end{align}
which should provide a complete set of eigenfunctions on our 
space-time. We will essentially recover the modes $\Upsilon^s_{k'}$  \eq{Upsilon} 
in the principal series of $SO(2,1)$.
 In the adapted $(t,\chi)$ coordinates and using \eq{BoxY-tchi}, this takes the form 
\begin{align}
 \Box_y \phi = 
   r^2  \big( -\del_\eta^2 - \tanh(\eta)\del_\eta
   + \tanh^2(\eta)\del^2_\chi \big) \phi = \l \phi \ .
\end{align}
To solve this equation,  we again make a separation ansatz
\begin{align}
 \phi = e^{i k \chi} \varphi_k(\eta)
\end{align}
which gives 
\begin{align}
- r^2 \big(\del_\eta^2 +\tanh(\eta)\del_\eta + k^2 \tanh^2(\eta) \big) \varphi_k
    = \l \varphi_k \ .
   \label{Box-eta-eq} 
\end{align}
Clearly for $\eta \to \pm \infty$ this reduces to the ordinary wave equation 
\begin{align}
   -r^2 \big(\del_\eta^2 + {\rm sign(\eta)} \del_\eta  +k^2  \big) \varphi_k
   \approx \l \varphi_k   , \qquad \eta \to \pm \infty
\end{align}
whose solutions for large  $k$ are exponentially damped plane waves,
\begin{align}
 \varphi_k^\pm(\eta) \to e^{\pm i k \eta - \frac 12 |\eta|},   \qquad \eta \to \pm \infty \ .
\label{asympt-solns-plane}
 \end{align}
We can bring the exact equation \eq{Box-eta-eq} into a more familiar form by
again substituting $u=\tanh(\eta) \in (-1,1)$ and
$
 f(u) = (1-u^2)^{1/4} h(u) 
$
to obtain  
\begin{align}
  (1 - u^2) h'' - 2 u h' + \left(-\left (k^2 + \tfrac 14 \right) + \frac{ k^2 + r^{-2} \l -\frac 14}{1-u^2} \right)
  h & = 0 \ .
\label{legendre-eq-2}
\end{align}
This has the same structure as \eq{legendre-eq}, replacing $-\l R^2 \to  k^2 + r^{-2} \l$.
It is hence solved again by associated Legendre functions of the
first and second kind
$\mathsf{P}_\nu^{\mu}$ and  $\mathsf{Q}_\nu^{\mu}$, as in section \ref{sec:harmonics-H2},
for
 \begin{align} 
 \qquad \nu(\nu+1) = - k^2- \frac 14 ~~~\mathrm{and} 
  \quad \mu = \pm i s,  \qquad s^2 =  k^2 + r^{-2} \l - \frac 14 .
 \label{Legendre-parameters-2}
\end{align}
Asymptotically oscillating solutions are obtained for $k^2 + \lambda/r^2 > \frac 14$
so that 
$\mu=\pm i s$ is purely imaginary,
\begin{align}
s = \sqrt{k^2 + \frac{\l}{ r^2} - \frac 14} \ > 0 ~.
 \label{order2:principal}
 \end{align}
A basis of solutions, as before, is given by
\begin{align}
\mathsf{P}^{is}_{\nu(k)}(u)
\end{align}
which form the unitary reps of $SO(2,1)$ of the principal series $P_s$.
The degree of the Legendre function can be taken to be
\begin{align}
  \nu(k) = -\frac 12 + i|k|
     \label{nu-k}
\end{align}
which should be compared with \eq{degree}. 
As expected, we obtain the same basis of modes as we did for $H^2$ in \eq{Upsilon},
\begin{align}
\boxed{\
\Upsilon^{s\pm}_k(\eta,\chi) := \frac 1{\sqrt{\cosh\eta}}\,
 e^{i k \chi} \LP^{\pm is}_{-\frac 12 + i |k|}(\tanh(\eta)) \qquad \mbox{for}
 \qquad s > 0, \ k\in\R \ .
 }
 \label{Upsilon-M}
\end{align}
To recap, above modes satisfy 
\begin{align}
\Box_y \Upsilon^{s\pm}_k &= r^2\big(s^2 - k^2 + \frac 14 \big) \Upsilon^{s\pm}_k , \nn\\
 (\Upsilon^s_k)^* &= \Upsilon^{-s}_{-k} \ .
 \label{Box-upsilon-M11}
\end{align}
These modes will be used to compute the path integral in section \ref{sec:path-integral}.

\paragraph{On-shell modes.}

Now we  identify the on-shell modes among the above harmonics,
which are the eigenmodes for $\l = m^2_\phi$. 
Then the eom \eq{eom-phi-class} has the following solutions
\begin{align}
 \cY^{\pm s}_k &= \frac 1{\sqrt{\cosh\eta}}e^{i k \chi} \LP^{\pm i s}_{\nu(k)}(\tanh(\eta))
 \end{align}
 where 
\begin{align} 
 \nu(k) &= - \frac 12 + i|k| ,  
 \qquad s^2 = k^2 + \frac{m_\phi^2}{r^2} -  \frac 14 \ .
 \label{nu-k-s}
\end{align}
These are the positive and negative energy eigenmodes, which form
principal series irreps.

\paragraph{Asymptotics and Bogoliubov coefficients.}

Since $s$ depends  now on $k$, the early and late time frequencies
depend on $k$. On-shell, we have
\begin{align}
 s=\omega_k:=\sqrt{ k^2 + \frac{m_\phi^2}{r^2} - \frac 14} \ .
 \label{omega-def}
\end{align}
The asymptotic expansion (\ref{P-asympt-neg}) and
(\ref{P-asympt-pos}) become
\begin{align}
\mathsf{P}^{ \pm i\omega_k}_{\nu(k)}\left(\tanh \eta \right)
~&\stackrel{\eta \to\infty}{\sim}~
\frac{ e^{ \mp i\omega_k  \eta}}{\Gamma\left(1 \mp i\omega_k\right)}~
 \label{P-asympt-neg-minkowski}
\end{align}
and
\begin{align}
\mathsf{P}^{ i\omega_k}_{\nu(k)}\left(\tanh \eta \right)
\stackrel{\eta \to -\infty}{\sim}
& - \frac{\sin\left(\nu\pi\right)\Gamma\left( i\omega_k \right)}{\pi} e^{ i\omega_k\eta}
\ + \ 
\frac {\Gamma(- i\omega_k)}{\Gamma\left(- i\omega_k-\nu\right)\Gamma\left( -i\omega_k+1+\nu\right)}
e^{- i\omega_k\eta}   ~.
 \label{P-asympt-pos-minkowski}
\end{align}
Therefore the modes $\cY^{+ s}_k \sim e^{i(k\chi -\omega_k \eta)}$ are negative energy modes in the far future
$\eta\to \infty$ (long after the BB),
if we consider $\eta$ as globally  oriented time coordinate, while $\cY^{- s}_k \sim e^{i(k\chi +\omega_k \eta)}$ are the positive energy modes.
In the far past $\eta\to -\infty$,  $\cY^{+ s}_k \sim \a_k e^{i(k\chi -\omega_k \eta)} + \b_k e^{i(k\chi +\omega_k \eta)}$ is then
a superposition of positive- and negative-energy modes. The transformation 
$\begin{pmatrix}  \a_k & \b_k \\
                 \b_k^* & \a_k^*
 \end{pmatrix}$ 
 is canonical i.e. it preserves the Poisson bracket.                                                
Comparing the coefficients in equations \eq{P-asympt-pos-minkowski} and  \eq{P-asympt-neg-minkowski},
we obtain the Bogoliubov coefficients:
\begin{align}
  \alpha_{k} &=  \frac     {\Gamma\left(1-i\omega_k \right)\Gamma(-i\omega_k)}{\Gamma\left(-i\omega_k-\nu\right)
    \Gamma\left(-i\omega_k+1+\nu\right)}
  =\frac{\Gamma(-i\omega_k)}{\Gamma(i\omega_k)\sin(i\pi\omega_k)}\frac\pi { \Gamma\left(-i\omega_k-\nu\right)\Gamma\left(-i\omega_k+1+\nu\right)}\nn
  \\
  \beta_{k} &=   -\frac{\sin\left(\nu\pi\right)\Gamma\left(1-i\omega_k \right) \Gamma\left(i\omega_k \right)}{\pi}~=
  ~-\frac{\sin \pi\nu}{\sin (i\pi\omega_k)} \ .
\end{align}
As a check, we can confirm that they satisfy $|\alpha_{k}|^2 - |\beta_{k}|^2 = 1$.
To do so, we notice that,
as long as $\mu= \pm i\omega_k$ 
is purely imaginary and Re$(\nu)=-\half$, $\half + \nu \pm \mu$ is purely imaginary,
and
\begin{align}
 \frac {1}{|\Gamma\left(\mu-\nu\right)\Gamma\left(\nu+\mu+1\right)|^2}  
  &= \frac 1{\pi^{2}} (\sin^2(\pi\nu) - \sin^2(\pi\mu)) \ .
\end{align}
We also have $|\sin\left(\mu\pi\right)|^2 = - \sin^2\left(\mu\pi\right) $ because
$\mu$ is purely imaginary, and $|\sin\left(\nu\pi\right)|^2 =  \sin^2\left(\nu\pi\right) $ because
the real part of $\nu$ is $\half$.  Then,
\begin{align}
  |\alpha_{k}|^2 - |\beta_{k}|^2  
   = \frac{\sin^2(\pi\nu) - \sin^2(\pi\mu)}{|\sin \pi\mu|^2}
~-~\frac{|\sin \pi \nu|^2}{|\sin \pi\mu|^2}
~=~   1 \ .
\end{align}
More explicitly, we have
\begin{align}
  |\beta_{k}|^2  &=    \frac{|\sin \pi \nu|^2}{|\sin \pi\mu|^2} =
  \frac{\cosh^2 \big(\pi\sqrt{k^2-\frac14}\big)}{\sinh^2(\pi\omega_k)} \ .
\end{align}
Using  the on-shell relation \eq{omega-def} we have
$\sqrt{k^2-\frac 14} \approx \omega_k$ in the relativistic regime, so that
\begin{align}
 |\beta_{k}|^2  &\approx \Big(\frac{e^{2\pi \omega_k} + 1}{e^{2\pi\omega_k}-1}\Big)^2
  \approx 1, \qquad
 |\alpha_{k}|^2 =  1+\b_k^2 \ \approx \ 2 \ .
\end{align}
This means that the Bogoliubov transformation is ``large'',
and strongly mixes the positive and negative energy modes.

\paragraph{Fuzzy wavefunctions.}

As discussed before, we define the fuzzy harmonics through the map in equation \eq{quantization-isometry}
with coefficients $c_s$ chosen so that \eq{isometry-condition} is satisfied,  
\begin{align}
\boxed{ \ \ 
 \{\hat\Upsilon^{\pm s}_k = \cQ(\Upsilon^{\pm s}_{k}) \} \ \ } \ .
 \label{Upsilon-hat}
\end{align}
These are the principal series modes in the unitary decomposition 
of $\End(\cH_n)$, cf. \eq{decompo}, and satisfy \eq{Laplacian-intertwiner}
\begin{align}
\Box_Y \hat\Upsilon^{\pm s}_k &= r^2\big(s^2 - k^2 + \frac 14 \big) \hat\Upsilon^{\pm s}_k , \nn\\
 (\hat\Upsilon^s_k)^\dagger &= \hat\Upsilon^{-s}_{-k} \ .
 \label{Box-upsilon-M11-fuzzy}
\end{align}
The equivalence via $\cQ$ implies that the matrix configurations have the same  properties as the classical ones,  and satisfy
 a unique time-evolution once the appropriate semi-classical boundary conditions  
 are imposed via $\cQ$. 
The local causality structure will be verified in the next section. 
In particular, the appearance of infinite time derivatives in a star product formulation is 
completely misleading in this respect, and the model with space-time noncommutativity 
has
perfectly nice and reasonable properties\footnote{Of course  non-commutativity does have significant implications.
Even though the correspondence defined via $\cQ$ is appropriate at low energies, it is quite misleading at high energies,
where the fields acquire a string-like behavior \cite{Steinacker:2016nsc}. This also implies that quantum effects in  interacting theories typically exhibit a strong non-locality known as UV/IR mixing.}.

\section{Fluctuations and path integral quantization}
\label{sec:path-integral}

The quantization of a matrix model is naturally  defined via a path integral,
which amounts to integrating over all matrices in $\End(\cH_n)$.
On the above background $\cM^{1,1}$, we can expand $\End(\cH_n)$ in the basis
$\hat\cY^{\pm s}_k$ of $SO(2,1)$ principal series modes \eq{Upsilon-hat},
\begin{align}
 \hat\phi = \int ds dk  \big(\phi^+_{s,k} \hat\Upsilon^{s+}_k
  + \phi^-_{s,k} \hat\Upsilon^{s-}_k\big) \qquad \in  \End(\cH_n)
\end{align}
integrating over $s>0$ and $k\in\R$. In the semi-classical limit,
this reduces to
\begin{align}
 \phi(x) = \int ds dk \big(\phi^+_{s,k} \Upsilon^{s+}_k(x)
 + \phi^-_{s,k} \Upsilon^{s-}_k(x)\big) \ .
\end{align}
We can now define  correlation functions  in the angular momentum basis as
\begin{align}
\langle \hat\phi^\s_{sk} \hat\phi^{\s'}_{s'k'}\rangle 
&:= \frac 1Z\,\int D\phi\, \phi^\s_{sk} \phi^{\s'}_{s'k'} 
e^{i S_\varepsilon[\phi]}  
\end{align}
were $\s,\s' = \pm$ and
 $D\phi = \Pi d\phi_{sk}$ is the integral over all  modes, 
and the  $i\varepsilon$ prescription \eq{iepsilon-mass} is understood.
Using the correspondence between classical and fuzzy functions, we can 
associate to this a 2-point function in position space as follows
\begin{align}
\langle \hat\phi(x) \hat\phi(y)\rangle 
&:=  \sum_{s,s'; \s,\s' = \pm}\Upsilon^{s\s}_k(x) \Upsilon^{s'\s'}_{k'}(y) 
  \langle \hat\phi^\s_{sk} \hat\phi^{\s'}_{s'k'}\rangle \ .
\end{align}
Since we only consider the free theory, the fuzzy case is equivalent to the 
semi-classical version on classical space-time.
The only new ingredient inherited from the matrix model 
is a specific action and the $i\varepsilon$ prescription\footnote{Since $\phi$ can be considered as a
transverse (space-like) matrix of the underlying Yang-Mills matrix model \eq{bosonic-action}, this prescription boils
down to replacing the mass term as $m^2\to m^2 - i \varepsilon$.}
\eq{iepsilon-mass}.

Now consider the action in terms of the eigenmodes, which in the semi-classical case
has the form 
\begin{align}
 S_\varepsilon[\phi] &= \int\limits_{H_2} \omega\phi^*(\Box_y - m^2_\phi +i\varepsilon)\phi \nn\\
   &= R r\int\limits_{H_2}\cosh(\eta) \big(\phi^+_{s',k'} \Upsilon^{s'+}_{k'} 
   + \phi^-_{s',k'} \Upsilon^{s'-}_{k'}\big)^*
      (s^2 - k^2 + \frac 14 - \tilde m^2 +i\varepsilon)
   \big(\phi^+_{s,k} \Upsilon^{s+}_{k} 
   + \phi^-_{s,k} \Upsilon^{s-}_{k}\big)
   \label{action-expand-full}
\end{align}
where the $\Upsilon^{s\pm}_{k} = (\Upsilon^{s\mp}_{-k})^*$ are given in \eq{Upsilon-M},
 the eigenvalue of $\Box_y$ 
is $r^2 (s^2 - k^2 + \frac 14)$ \eq{Box-upsilon-M11-fuzzy} and
\begin{align}
 \tilde m^2 = \frac{m^2}{r^2} \ .
\end{align}
To evaluate the action, we need
\begin{align}
\int d\chi d\eta \cosh(\eta)(\Upsilon^{s'}_{k'})^*\Upsilon^{s}_{k} 
 &= \int_{-\infty}^\infty d\chi e^{i (k'-k)\chi} \int_{-1}^1 \frac{du}{1-u^2} 
 \LP^{-is'}_{\nu(k')} (u)  \LP^{is}_{\nu(k)} (u)  \nn\\
  &= (2\pi)\d(k-k') \int_{-1}^1 \frac{du}{1-u^2} 
  \LP^{-is'}_{\nu(k)} (u)  \LP^{i s}_{\nu(k)} (u)   \nn\\
  &= (2\pi)\d(k-k') \Big(a(k,s)  \d(s+s') + b(k,s) \d(s-s')\Big)
 \label{orthog-M11}
\end{align}
cf. \eq{orth-H2-explicit}
using the orthogonality relations \eq{Ipm-orth}, where 
\begin{align}
 a(k,s) 
 &= \frac{2\pi}{\G(\frac 12 + i |k|-is)\G(\frac 12- i|k|-is)}\frac{\cosh(\pi k)}{s\sinh(\pi s)}
 = a(k,-s)^* \nn\\
 b(k,s)
 &=  2\frac{\sinh(\pi s)}{s}\Big(1 + \frac{\cosh^2(\pi k)}{\sinh^2(\pi s)}\Big)
 = b(k,-s) \ .
\end{align}
Note that half of the terms in \eq{action-expand-full} will drop out since $s,s' > 0$.
We thus obtain 
\begin{align}
  S_\varepsilon[\phi] 
 &= 2\pi R r \int ds dk ds' dk' \d(k-k') \d(s-s') 
   (s^2 - k^2 + \frac 14 - \tilde m^2 +i\varepsilon) \nn\\
 &\qquad  \begin{pmatrix}
           (\phi^+_{s',k'})^*,(\phi^-_{s',k'})^*
          \end{pmatrix}
          \begin{pmatrix}
           b(k,s) & a(k,-s)  \\
           a(k,s) & b(k,-s)
          \end{pmatrix}
\begin{pmatrix}
 \phi^+_{s,k} \\
 \phi^-_{s,k} 
\end{pmatrix} \ .
\end{align}
Inverting the $2\times 2$ matrix,  the propagator in ``momentum space'' is 
\begin{align}
\left \langle\begin{pmatrix}
 \phi^+_{s,k} \\
 \phi^-_{s,k} 
\end{pmatrix}
\left ((\phi^+_{s',k'})^*,(\phi^-_{s',k'})^* \right ) \right \rangle 
 &= \frac{1}{\pi R r } \d(k-k') \d(s-s')
    \frac 1{s^2 - k^2 + \frac 14 - \tilde m^2 +i\varepsilon} \nn\\
 &\qquad  \frac{s^2}{\cosh (2 \pi s)+\cosh (2 \pi k)}
          \begin{pmatrix}
           b(k,-s) & -a(k,-s)  \\
           -a(k,s) & b(k,s)
          \end{pmatrix} 
\end{align}
using 
\begin{align}
 \det\begin{pmatrix}
           b(k,s) & a(k,-s)  \\
           a(k,s) & b(k,-s)
          \end{pmatrix}
 = \frac{2}{s^2} (\cosh (2 \pi  k)+\cosh (2 \pi  s)) \ .
\end{align}

\subsection{Propagator in position space}

In the $\eta\chi$ space-time coordinates of $\cM^{1,1}$, the propagator takes the form 
\begin{align}
\langle \hat\phi(\eta,\chi) \hat\phi(\eta',\chi')^*\rangle 
 &= \int ds dk ds' dk'\sum_{\s,\s'=\pm}
 \Upsilon^{s\s}_k(\eta,\chi) 
  \langle \hat\phi^\s_{sk} (\hat\phi^{\s'}_{s'k'})^*\rangle  
  \Upsilon^{s'\s'}_{k'}(\eta',\chi')^* \ .
  \label{propagator-def}
\end{align}
We can evaluate this  explicitly in the late-time regime $\eta\to\infty$ 
using the asymptotic form  \eq{P-asympt-neg-minkowski}, which gives  
\begin{align}
 \Upsilon^{s\pm}_{k} &= 
 \frac{e^{i k \chi}}{\sqrt{\cosh \eta}}  \mathsf{P}^{\mp is}_{\nu(k)}\left(\tanh \eta \right)  
~~\stackrel{\eta \to \infty}{\sim}~~
\frac{e^{i (k \chi \pm s\eta)}}{\Gamma(1 \mp is)\sqrt{\cosh \eta}}  \ .
\end{align}
Thus 
\begin{align}
 \langle \hat\phi(\eta,\chi) \hat\phi(\eta',\chi')^*\rangle 
 &\sim
 \frac{2}{2\pi R r \sqrt{\cosh\eta\cosh\eta'}}
 \int \!\! ds dk \frac{e^{i k (\chi-\chi')} }
    {s^2 - k^2 + \frac 14 - \tilde m^2 +i\varepsilon}
    \frac{s^2}{\cosh (2 \pi s)+\cosh (2 \pi k)} \nn\\
 &\qquad \big(\frac{e^{-i s\eta}}{\Gamma(1 - is)},\frac{e^{i s\eta}}{\Gamma(1 + is)}\big)
  \begin{pmatrix}
           b(k,-s) & -a(k,-s)  \\
           -a(k,s) & b(k,s)
          \end{pmatrix} 
          \begin{pmatrix}
           \frac{e^{i s\eta'}}{\Gamma(1 + is)} \\ \frac{e^{-i s\eta'}}{\Gamma(1 - is)}
          \end{pmatrix} \nn\\
 &=  \frac{1}{2\pi R r \sqrt{\cosh(\eta)\cosh(\eta')}}
 \int ds dk \frac{e^{i k (\chi-\chi')} }
    {(s^2 - k^2 + \frac 14 - \tilde m^2 +i\varepsilon)} \nn\\
 &\qquad \Big [\frac{\cos (s (\eta-\eta'))}{\pi }
 + \frac 1{2\pi^3}\cosh (\pi k) s\sinh (\pi s)  \nn\\
 &\qquad \big(e^{ -i s (\eta+\eta')} \Gamma (i s)^2 
 \Gamma(i k-i s+\frac{1}{2}) \Gamma(-i k-i s+\frac{1}{2}) \nn\\
 &\qquad +e^{i s (\eta+\eta')}\Gamma (-i s)^2 \Gamma(i k+i s+\frac{1}{2}) \Gamma(-i k+i s+\frac{1}{2})
 \big)\Big ]  \nn\\
 %
 %
&=:  \langle \hat\phi(\eta,\chi) \hat\phi(\eta',\chi')^*\rangle_0
   +  \langle \hat\phi(\eta,\chi) \hat\phi(\eta',\chi')^*\rangle_{op} \ .
\end{align}
At late times $\eta, \eta' \to \infty$, the second term is rapidly 
oscillating and hence suppressed. 
Therefore the first term is the leading contribution in the late time regime.

\paragraph{Late time propagator for $\eta', \eta \to\infty$.}

Consider first the late time propagator
\begin{align}
 \langle \hat\phi(\eta,\chi) \hat\phi(\eta',\chi')^*\rangle_0 
 &= c_\eta
 \int_{-\infty}^\infty ds \int_{\infty}^\infty dk 
 \frac{e^{i (k (\chi-\chi') + s(\eta-\eta'))}}
    {s^2 - k^2 + \frac 14 - \tilde m^2 +i\varepsilon} \ .
    \label{samesheet-correl}
\end{align}
The  pre-factor
\begin{align}
 c_\eta := \frac 1{4\pi^4 R r \sqrt{\cosh \eta\cosh \eta'}}
\end{align} 
reflects the non-canonical normalization, which can be traced to the exponential damping
behavior in \eq{asympt-solns-plane}.
Apart from this normalization, we  recover precisely the Feynman propagator on flat
1+1-dimensional space-time at zero temperature, including the 
appropriate $i\varepsilon$ prescription which ensures local causality.

Notice that the formula applies equally in the opposite limit $\eta, \eta' \to -\infty$. 
Since the eigenmodes stretch continuously across the singularity at $\eta=0$, the
parameter $\eta$ is expected to indicate the physical time evolution 
on both sides of the Big Bounce, so that the arrow of time points
inwards (towards the BB) for $\eta<0$. 
This strongly suggests to interpret the singularity as ``Big Bounce''.
A more profound justification e.g. via entropic considerations 
is beyond the scope of this paper.

\paragraph{Non-local contribution for large $\eta\approx -\eta'\to\infty$.}

To evaluate \eq{propagator-def} in a limit where $\eta \rightarrow \infty$ but $\eta' \rightarrow -\infty$,
we make use of the asymptotic form \eq{P-asympt-pos-minkowski} and the Bogoliubov coefficients:
\begin{align}
 \langle \hat\phi(\eta,\chi) \hat\phi(\eta',\chi')^*\rangle 
 &\sim
 \frac{2}{2\pi R r \sqrt{\cosh\eta\cosh\eta'}}
 \int \!\! ds dk \frac{e^{i k (\chi-\chi')} }
    {s^2 - k^2 + \frac 14 - \tilde m^2 +i\varepsilon}
    \frac{s^2}{\cosh (2 \pi s)+\cosh (2 \pi k)} \nn\\
     & \big(\frac{e^{-i s\eta}}{\Gamma(1 - is)},\frac{e^{i s\eta}}{\Gamma(1 + is)}\big)
  \begin{pmatrix}
           b(k,-s) & -a(k,-s)  \\
           -a(k,s) & b(k,s)
  \end{pmatrix}
    \begin{pmatrix}
           \a^*(k,s) & \b^*(k,s)  \\
           \b(k,s) & \a(k,s)
          \end{pmatrix} 
          \begin{pmatrix}
           \frac{e^{i s\eta'}}{\Gamma(1 + is)} \\ \frac{e^{-i s\eta'}}{\Gamma(1 - is)}
          \end{pmatrix} \ ,\nn
\end{align} 
where 
\begin{align}
  \a(k,s) &:=  \frac     {\Gamma\left(1-is \right)\Gamma(-is)}{\Gamma\left(-is-\nu\right)
    \Gamma\left(-is+1+\nu\right)}
  =-i\frac{\Gamma(-i s)}{\Gamma(i s)\sinh(\pi s)}\frac\pi { \Gamma\left(-i s-\nu\right)\Gamma\left(-i s+1+\nu\right)}\nn
  \\
  \b(k,s) &:=   -\frac{\sin\left(\nu\pi\right)\Gamma\left(1-i s \right) \Gamma\left(is \right)}{\pi}~=
  ~i\frac{\cosh \pi k}{\sinh (\pi s)} = -\b^*(k,s) \ .
  \label{alpha-beta-bogol-off-shell}
\end{align}
Note that
\begin{align}
 a(k,s) = a(k,-s)^*
 &= 
 \frac 2 {-is}  \frac{\Gamma(is)}{\Gamma(-is)}\cosh (\pi k)~ \a(k,s)~,
 \end{align}
and
\begin{align}
 b(k,s) =  b(k,-s) 
 &= 
  \frac 2 {-is}\frac{\sinh^2(\pi s)+\cosh^2(\pi k)}{  \cosh (\pi k)}~ \b(k,s)
\ ,
 \end{align}
and, as before,
\begin{align}
\det    \begin{pmatrix}
           b(k,-s) & -a(k,-s)  \\
           -a(k,s) & b(k,s)
          \end{pmatrix} 
&= \frac 4 {s^2} \left ( \sinh^2(\pi s) + \cosh^2(\pi k) \right ) ~=~
\frac 2 {s^2} \left ( \sinh(2\pi s) + \cosh(2\pi k)  \right ) \nn =: D \ ,
\end{align}
where we defined a useful quantity $D$:
\begin{align}
  |\a(s,k)|^2 = 1 + |\b(s,k)|^2  & =  
  \frac{ Ds^2}{4 \sinh^2(\pi s)} 
\nn \\ a(k,s) = a(k,-s)^*
 &= -
 \frac {2\cosh (\pi k)} {is}  \frac{\Gamma(is)}{\Gamma(-is)}~ \a(k,s) 
  \nn\\
 b(k,s) =  b(k,-s) 
 &=  D \frac {is} {2  \cosh(\pi k)}~ \b(k,s) ~=~ \frac {is} {2  \sinh(\pi s)} ~D
\ ,
\end{align}
This allows us to evaluate:
\begin{align}
-a(k,s)\a^*(k,s) + b(k,s) \b(k,s) 
& = -\frac{iDs}{2} \frac{ \cosh(\pi k)}{\sinh^2(\pi s)}
\left ( \frac{\Gamma(is)}{\Gamma(-is)} + 1\right ) \ ,
\end{align}
which allow us to identify in the combination
\begin{align}
   \big(\frac{e^{-i s\eta}}{\Gamma(1 - is)},\frac{e^{i s\eta}}{\Gamma(1 + is)}\big)
  \begin{pmatrix}
           b(k,-s) & -a(k,-s)  \\
           -a(k,s) & b(k,s)
  \end{pmatrix}
    \begin{pmatrix}
           \a^*(k,s) & \b^*(k,s)  \\
           \b(k,s) & \a(k,s)
          \end{pmatrix} 
          \begin{pmatrix}
           \frac{e^{i s\eta'}}{\Gamma(1 + is)} \\ \frac{e^{-i s\eta'}}{\Gamma(1 - is)}
          \end{pmatrix} \ ,\nn
\end{align}
terms  that do not oscillate rapidly in the limit considered.  One of these is
\begin{align}
& -  e^{is(\eta+\eta')}
\left [ -a(k,s)\a^*(k,s) + b(k,s) \b(k,s)  \right ] \frac{1}{\Gamma(1+is)} \frac{1}{\Gamma(1+is)} =
  \nn \\&  -e^{is(\eta+\eta')}\frac{D}{2} \frac{\cosh(\pi k)}{\pi^2} 
  \left [
      {\Gamma(is)}~  
+ {\Gamma(-is)}~  \right ]{\Gamma(1-is)} 
  \end{align}
and the other is its complex conjugate.
The leading part of the propagator is therefore
\begin{align}
 \langle \hat\phi(\eta,\chi) \hat\phi(\eta',\chi')^*\rangle 
 &\sim
  -\frac{1}{\pi^3 R r \sqrt{\cosh\eta\cosh\eta'}}
 \int\limits_{s>0} \!\! ds dk \frac{e^{i k (\chi-\chi')}  \cosh(\pi k)}
    {s^2 - k^2 + \frac 14 - \tilde m^2 +i\varepsilon}
    \nn\\ 
    &\left [
    e^{is(\eta+\eta')}
      \Gamma(1-is)\left ({\Gamma(is)}~
+ {\Gamma(-is)} \right )~  + ~ \mathrm{c.c.}~~ \right ] \nn\\
 &= 
 -\frac{1}{\pi^3 R r \sqrt{\cosh\eta\cosh\eta'}}
 \int \!\! ds dk \frac{e^{i k (\chi-\chi')}  e^{is(\eta+\eta')}}
    {s^2 - k^2 + \frac 14 - \tilde m^2 +i\varepsilon} \cosh(\pi k)\Phi(s)
    \label{oppsheet-correl}
\end{align} 
(the integral is over $s \in \R$ in the last expression)
for $\eta \rightarrow \infty$ but $\eta' \rightarrow -\infty$. Here
\begin{align}
  \Phi(s) := \Gamma(1-is)\left ({\Gamma(is)}~  + {\Gamma(-is)} \right ) \ =
  \frac 1 {\sinh\pi s}  \left ( 1 + \frac{\Gamma(-is)}{\Gamma(is) }\right )
\end{align}
is a regular function in $s\in\R$ which decays exponentially 
for large $s$:
 \begin{align}   
   |\Phi(s)| \sim e^{-\pi |s|}, \qquad s\to \pm\infty \ .
 \end{align}
 However, the expression in equation (\ref{oppsheet-correl})
 is pathological due to the $\cosh(\pi k)$ factor, which leads to a UV divergence of the
space-like momentum $k$.
This divergence can be cured by smearing the correlation functions by a space-like Gaussian
$\psi_{\chi_0}(\chi) = \frac 1{\sqrt{\s \pi}\,}e^{- (\chi-\chi_0)^2/2\s^2}$
with width $\s$:
\begin{align}
\langle \hat\phi(\eta,\chi_0) \hat\phi(\eta',\chi'_0)^*\rangle_\s :=
 \int d\chi d\chi'\psi_{\chi_0}(\chi)\langle \hat\phi(\eta,\chi) \hat\phi(\eta',\chi')^*\rangle \psi_{\chi'_0}(\chi') \ .
\end{align}
Noting that
 $\int d\chi  e^{- (\chi-\chi_0)^2/\s^2} e^{i k \chi}
  =  e^{-\frac{\s k^2}{4}} e^{i k \chi_0}$
this space-like UV divergence then disappears:
\begin{align}
\langle \hat\phi(\eta,\chi) \hat\phi(\eta',\chi')^*\rangle_\s
  &= -\frac{1}{\pi^3 R r  \sqrt{\cosh\eta\cosh\eta'}}
 \int \!\! ds dk \frac{ e^{is(\eta+\eta')}e^{i k (\chi_0-\chi_0')} \cosh(\pi k)e^{-\frac{\s k^2}{2}} }
    {s^2 - k^2 + \frac 14 - \tilde m^2 +i\varepsilon} \Phi(s) \ .
\end{align}
Now the integrals are  well-defined. Due to their oscillatory 
behavior,  the correlators  are peaked at $\eta \approx -\eta'$
and $\chi_0 \approx \chi_0'$ and strongly suppressed otherwise.
We therefore obtain a non-trivial correlation between the fields 
before and after the Big Bounce, for  points on the in-  and out sheets
which coincide in target space.
This result will find a natural interpretation in terms of string states, as discussed below.

It is remarkable that the correlations between smeared wave-packets between
the in-and out-sheets are   perfectly
well defined, while the point-like propagators are not\footnote{Note that the same result holds also
in the matrix case, since the free theory in the commutative and matrix framework are
identical, related by a Weyl-type quantization map. However due to UV/IR mixing or
the uncertainty relation on NC spaces, $k\to\infty$ inevitably entails $s\to 0$. 
Therefore the UV divergence in $k$
would disappear e.g. on compact space-times, such as a cyclic cosmologies. The 
ramifications in the presence of interactions are unclear.}.
This indicates that the Bogoliubov transformation
relating the in- and out vacua on the two sheets 
strongly modifies the UV structure of the  modes, which is also 
manifest in \eq{alpha-beta-bogol-off-shell}.
The physical significance of this observation is not clear, and
deserves further investigations.

\subsection{Further remarks}

In the noncommutative or matrix setting, the above calculation goes through  
for the free theory, because the spectrum of $\Box$
coincides with the commutative case,
and the eigenmodes are in one-to-one correspondence via $\cQ$.
In the presence of interactions,
only the IR modes  behave as in the commutative theory, while the UV sector 
is better described by non-local string modes $|x\rangle\langle y|$
\cite{Steinacker:2016nsc,Steinacker:2022kji}; these also
provide a geometrical understanding 
 of the spectrum of $\Box$.
In noncommutative field theory, such
non-local string modes span the extreme UV sector of the theory with eigenvalues
$\Box \sim |x-y|^2 + \L_{NC}^2$ far above the scale of noncommutativity $\L_{NC}$,
and they are responsible for UV/IR mixing.
 
Due to the 2-sheeted structure of the present $\cM^{1,1}$ brane,
there are  in particuar string modes of
the structure 
\begin{align}
 |x\rangle_+\langle y|_- \qquad \in \End(\cH_n) \ 
\end{align}
which connect the pre-BB and post-BB sheets;
here $|x\rangle_+$ is a coherent state  on the upper (post-BB) sheet and
$|y\rangle_-$ is a coherent state  on the lower (pre-BB) sheet.
From the point of view of either sheet, they behave like point-like objects 
which are charged under $U(1)$.
In particular, the  antipodal points on the opposite sheets of $\cM^{1,1}$ coincide
in target space, so tha the corresponding string modes 
have only  ``intermediate'' energy of the order  $\L_{NC}$.
These modes appear to be responsible for the observed
correlation for $\eta + \eta' \approx 0$, which are
 non-local from the intrinsic brane point of view,
 but local in target space.
A similar phenomenon can be seen for the squashed fuzzy sphere, cf. \cite{Andronache:2015sxa}.

Although the string states are typically UV states, 
they are important in the loops, and mediate 
long-distance interactions \cite{Steinacker:2016nsc}. 
In particular, the inter-brane string states 
connecting the two branes  will 
lead to gravity-like interactions between the pre-BB and post-BB branes at one loop.
This effect is on top of the correlations observed in the previous  section, 
which arise in the free theory. 
The same effects will  apply in the more realistic
3+1-dimensional cosmological solution \cite{Sperling:2019xar}.
It is therefore conceivable that physically significant 
correlations and  interactions exist between the pre-BB and post-BB branes.
Such effects would  be very intriguing,
but they arise only for the specific embedding structure
of the coincident branes in target space under consideration.

Finally, there is a subtlety in the signature of the effective metric, which is somewhat hidden in our analysis.
The  effective metric on noncommutative branes in Yang-Mills matrix models
has the structure $G^{\mu\nu} =  \theta^{\mu\mu'}\theta^{\nu\nu'}\eta_{\mu'\nu'}$ \cite{Steinacker:2010rh},
which is closely related to the open string metric \cite{Seiberg:1999vs}. In the presence
of time-like noncommutativity,  the anti-symmetric structure
of the Poisson tensor $\theta^{\mu\mu'}$ implies a flip of the causality structure, which in
1+1 dimensions amounts to a flip of the space- and time-like directions. In the
scalar field theory under consideration, this can be accommodated simply by an appropriate
choice of overall sign. This phenomenon disappears  on the covariant quantum space-times
discussed in \cite{Sperling:2019xar,Steinacker:2017vqw}, which have a very similar 3+1-dimensional structure as the present background.
Since the $i\varepsilon$ regularization of the matrix model is independent of the background,
the conclusions of the present paper can be extended straightforwardly to these 3+1-dimensional
backgrounds \cite{Battista:2022hqn}.

\section{Conclusion}
\label{sec:discussion}

In this paper, we have demonstrated some new and remarkable features of 
field theory on Lorentzian noncommutative space-time in matrix models.
In particular, we have shown that a suitable regularization of the 
Lorentzian (oscillatory) matrix path integral  leads to the usual
$i\varepsilon$ prescription for the emergent local quantum field theory, even on a curved background.
We obtained the propagator on a non-trivial 1+1-dimensional FLRW-type background
by computing the ``matrix'' path integral \eq{path-integral-eps},
which is seen to reduce locally to the standard Feynman propagator. 

This result demonstrates that the framework of
Yang-Mills matrix models, including notably the IKKT model, can indeed give rise to
a physically meaningful time evolution, even though there is no a priori time in the matrix model.
This should be contrasted to models of matrix quantum mechanics such as the BFSS model
\cite{Banks:1996vh,deWit:1988wri},
which are defined in terms of an a priori notion of time.
Even though we consider only a simple, free toy model in 1+1 dimensions,
the result clearly extends to the interacting case.
However then UV/IR mixing arises due to non-local string states,
so that a sufficiently local theory should be expected 
only for the maximally supersymmetric IKKT model.

From a physics perspective, perhaps the most interesting conclusion is that
the modes and the propagator naturally extend across the Big Bounce.
It is therefore possible to study questions
such as the propagation of physical modes across the BB, in a well-defined framework of
quantum geometry provided by the matrix model.
For the particular space-time solution under consideration, 
we also observe an intriguing correlation between
 the pre-BB and post-BB physics, which is attributed to the coincidence of the 
 pre-and post-BB sheets in target space.
 All these  results generalize  to an analogous 3+1-dimensional solution
 \cite{Battista:2022hqn}.
 However, we leave a more detailed investigation
 of these and other physical aspects to future work.

\paragraph*{Acknowledgements.}

Useful discussions with Robert Brandenberger at the EISA summer institute
are gratefully acknowledged, as well as
a related collaboration with Emmanuele Battista. 
The work of HS was supported by the Austrian Science Fund (FWF) grants P 28590 and  P 32086.
The work of JLK was supported by the Natural Sciences and Engineering Council
of Canada (NSERC), grant SAPIN-2016-00032.

\section{Appendix}

\subsection{Unitary representations of \texorpdfstring{$\mso(2,1)$}{so(2,1)}}
\label{sec:unitary-reps}

Unitary representations are characterized by hermitian generators
\begin{align}
 K_a^\dagger = K_a, \quad a=0,1,2 .
\end{align}
acting on the weight basis as follows
\begin{align}
K_3|j,m\rangle&= m|j,m\rangle, \nn\\
K_+|j,m\rangle&= a_{m+1}|j,m+1\rangle, \nn\\
K_-|j,m\rangle&= a_m|j,m-1\rangle,
\label{expabsalg}
\end{align}
where
\begin{align}
a_m=\sqrt{m(m-1)- j(j-1)}.
\label{norma}
\end{align}
There are different classes of unitary irreps of $\mso(2,1)$ (see e.g. 
\cite{Bargman-unitary}):
 
$\bullet$ The discrete series 
\begin{align}
D_j^{+}, \quad j \in \N_{>0}: \qquad  \cH_j &= \{|j,m\rangle; m = j,j+1,\cdots ;m \in \N\; \},  \nn\\
D_j^-, \quad j \in \N_{>0}: \qquad \cH_j  &= \{|j,m\rangle; m = -j,-j-1,\cdots ;-m \in \N\;\},
\label{discrete}
\end{align}
characterized by a Casimir $C^{(2)}=j(j-1)\geq 0$.
 These are either lowest or highest weight irreps, which 
 correspond to square-integrable positive or negative energy
 wavefunctions on $AdS^2$.
The states which span the positive energy (lowest weight) irreps
are obtained by acting with $K^+$ on the lowest-weight state, and conversely
for the negative energy (highest weight) irreps.

$\bullet$ The principal continuous series 
\begin{align}
P_s, \quad s \in \R,\quad 0<s<\infty,\quad  j = \frac 12 + i s:
\qquad \cH_j = \{|j,m\rangle; m=0,\pm 1, ...; m  \in \Z \} \ ,
\label{principal}
\end{align}
is labeled by a real number $s$ and has $C^{(2)}= -\left( s^2 + \frac 14\right)<-1/4$.
 These correspond to wavefunctions on the 
hyperboloid $H^2$, which is the space 
of interest this paper.
Note that the eigenvalues of $K^3$ are real, but $j$ is complex.

$\bullet$ The complementary series
\begin{align}
P_j^c, \quad  1/2 < j < 1,\qquad j \in \R :
\qquad \cH_j = \{|j,m\rangle; m=0,\pm 1, ...; m  \in \Z \}
\label{complementary}
\end{align}
with $-1/4<C^{(2)}<0$.

\subsection{Normalization of the Legendre functions}

In \cite{bielski2013orthogonality}, the following formula was given
\begin{align}
 I(s,s') &= \int_{-1}^1 \LP^{i s}_\nu(t) \LP^{i s'}_\nu(t) \frac{1}{1-t^2}dt \nn\\
  &= -\frac{2\G(is) \G(-is)}{\G(1+\nu-is)\G(-\nu-is)}\sin(\pi\nu)\d(s-s')\nn\\
  &\quad +\Big(\frac{\pi}{\G(1-is)\G(1+is)} + \frac{\sin^2(\pi\nu)\G(is)\G(-is)}{\pi}\nn\\
  &\quad
  + \frac{\pi\G(is)\G(-is)}{\G(1+\nu-is)\G(-\nu-is)\G(1+\nu+is)\G(-\nu+is)}\Big) \d(s+s') \nn\\
  %
 %
 &= -\frac{2\pi}{\G(1+\nu-is)\G(-\nu-is)}\frac{\sin(\pi\nu)}{s\sinh(\pi s)}
   \d(s-s')  \nn\\
  &\quad +2\Big( \frac{\sinh(\pi s)}{s}
  + \frac{\sin^2(\pi\nu)}{s\sinh(\pi s)}\Big) \d(s+s') 
  \label{Ipm-orth}
\end{align}
using the standard identities
\begin{align}
 \G(is)\G(-is) &= \frac{\pi}{s\sinh(\pi s)} \nn\\
 \G(1+is)\G(1-is) &= \frac{\pi s}{\sinh(\pi s)} \nn\\
 \G(z)\G(1-z) &= \frac{\pi}{\sin(\pi z)}
\end{align}
and
\begin{align}
 \sin(\pi(-\nu-is))\sin(\pi(-\nu+is))
  &= \sin^2(\pi\nu) + \sinh^2(\pi s) \ .
\end{align}
For $\nu = -\frac 12 + i |k|$, this is
\begin{align}
 I(s,s') &= \frac{2\pi}{\G(\frac 12 + i(|k| -s))\G(\frac 12 - i(|k| +s))}
 \frac{\cosh(\pi k)}{s\sinh(\pi s)}
   \d(s-s')  \nn\\
  &\quad +2\Big( \frac{\sinh(\pi s)}{s}
  + \frac{\cosh^2(\pi k)}{s\sinh(\pi s)}\Big) \d(s+s')
\end{align}
noting that 
\begin{align}
 \sin^2(\pi(-\frac 12 + i |k|)) &= \cosh^2(\pi k) \ .
\end{align}

%

%
%

\bibliographystyle{JHEP}
\bibliography{papers}

\newpage%

\end{document}